\documentclass[reprint,
superscriptaddress,
nofootinbib,
 aps,
prd,
]{revtex4-2}

\usepackage{graphicx}
\usepackage{xcolor}
\usepackage{amsmath}
\usepackage{amssymb}
\usepackage{bm}
\usepackage{mleftright}
\usepackage[ISO]{diffcoeff}
\usepackage{siunitx}
\usepackage{natbib}
\usepackage{hyperref}
\usepackage[capitalise]{cleveref}

\makeatletter
\input{aas_macros.sty}
\let\jnl@style=\relax
\makeatother

\crefname{section}{section}{sections}

\DeclareSIUnit{\yr}{yr}
\DeclareSIUnit{\Gyr}{\giga\yr}
\DeclareSIUnit{\pc}{pc}
\DeclareSIUnit{\kpc}{\kilo\pc}
\DeclareSIUnit{\Mpc}{\mega\pc}
\DeclareSIUnit{\Msun}{\text{\ensuremath{M_\odot}}}
\DeclareSIUnit{\hHubble}{\text{\ensuremath{h}}}
\newcommand*{\rh}{\ensuremath{r_{1/2}}}

\begin{document}

\title{Updated bounds on ultra-light dark matter from the tiniest galaxies}

\author{Simon May}
\email{simon.may@pitp.ca}
\affiliation{Perimeter Institute for Theoretical Physics, 31 Caroline Street N., Waterloo, Ontario, N2L 2Y5, Canada}

\author{Neal Dalal}
\email{ndalal@pitp.ca}
\affiliation{Perimeter Institute for Theoretical Physics, 31 Caroline Street N., Waterloo, Ontario, N2L 2Y5, Canada}

\author{Andrey Kravtsov}
\affiliation{Department of Astronomy \& Astrophysics, The University of Chicago, Chicago, IL 60637, USA}
\affiliation{Kavli Institute for Cosmological Physics, The University of Chicago, Chicago, IL 60637, USA}
\affiliation{Enrico Fermi Institute, The University of Chicago, Chicago, IL 60637, USA}


\begin{abstract}
The particle mass of dark matter (DM) was previously constrained using kinematics of ultra-faint dwarf galaxies to $m > \SI{3e-19}{\eV}$. This constraint, which excludes the ``fuzzy'' range of ultra-light dark matter from comprising all of the DM, relies on an estimate of the heating rate from fuzzy dark matter (FDM) wave interference using linear perturbation theory. Here, we compare the results of this perturbative calculation to full Schrödinger–Poisson simulations of the evolution of star particles in FDM halos. This comparison confirms theoretical expectations that FDM heating is \emph{stronger} in fully nonlinear simulations due to the formation of a dense central soliton whose fluctuations enhance gravitational perturbations, and that bounds on the DM particle mass using this perturbative method are indeed conservative.
We also show that these bounds are not affected by possible tidal stripping, since for dwarf satellites like Segue 1, the tidal radius is much larger than the observed size of the galaxy. We further show that the constraints on the mass cannot be evaded by invoking DM self-interactions, due to constraints on the self-interaction from large-scale structure. Lastly, we show that if the recently discovered system Ursa Major III/UNIONS I is a galaxy, the observed properties of this object strengthen the lower bound on the DM mass by over an order of magnitude, to $m > \SI{8e-18}{\eV}$, at \SI{95}{\percent} confidence. This constraint could further be strengthened considerably by more precise measurements of the size and velocity dispersion of this and other similar galaxies, and by using full Schrödinger–Poisson simulations.
\end{abstract}

\maketitle

\section{Introduction}
\label{sec:intro}

Axion-like particles are appealing dark matter (DM) candidates with allowed masses spanning many orders of magnitude \cite{Arvanitaki2010}. If DM is composed of ultra-light particles, with masses far below an \si{\eV}, then their occupation number far exceeds unity, allowing us to treat DM in this regime as a coherent classical field exhibiting wave phenomena like interference \cite{Hui2021}. In this regime, an interference effect analogous to the Hanbury~Brown–\allowbreak Twiss effect in optics \cite{HanburyBrown1954, Dalal2024} leads to $\mathcal{O}(1)$ fluctuations in the energy density of DM in virialized structures like DM halos. These density fluctuations have coherence lengths of order $\hbar/(m\sigma_v)$, and coherence times of order $\hbar/(m\sigma_v^2)$, where $m$ is the DM particle mass, and $\sigma_v$ is the velocity dispersion of DM.

These density fluctuations source gravitational perturbations and can thus perturb motions of stars embedded in the DM halo potential. Since stars are typically born from cold gas deep inside the halo, they initially have velocity dispersions far below the DM velocity dispersion. The gravitational fluctuations from DM wave interference can act to dynamically heat the motions of stars over time. This heating effect grows with increasing de Broglie wavelength, and is thus strongest in the smallest galaxies with the lowest $\sigma_v$. For this reason, ultra-faint dwarf galaxies with sizes $\rh \lesssim \SI{50}{\pc}$, velocity dispersions $\sigma_v \lesssim \SI{5}{\km\per\s}$, and mass-to-light ratios so large that stellar self-gravity is utterly negligible \cite{Simon2019}, provide extremely sensitive probes of ultra-light dark matter (ULDM).

Two of the smallest ultra-faint dwarf galaxies, Segue~1 \cite{Simon2011} and Segue~2 \cite{Kirby2013}, have been used to constrain the DM particle mass to $m > \SI{3e-19}{\eV}$ (\citep{Dalal2022}, hereafter DK22). This constraint excludes the entire ``fuzzy'' range of ULDM, $\SI{e-22}{\eV} < m < \SI{e-20}{\eV}$ \cite{Hu:2000ke, Hui2021} from being all of the DM, and is completely insensitive to uncertainties in baryonic physics, such as supernova feedback, the galaxy–halo connection, etc., since it does not rely on predictions of the central DM density in these galaxies, but instead uses only the \emph{observed} DM densities to compute the gravitational perturbations arising from the $\delta\rho \approx \bar{\rho}$ density fluctuations that are unavoidable in this ultra-light regime.

In this paper, we revisit the bounds on ultra-light dark matter provided by the smallest galaxies. There are several reasons to do so now. First, the constraints on the DM mass $m$ determined by DK22 relied on an approximate treatment of DM wave interference using linear perturbation theory. DK22 argued that their method should provide a conservative bound on $m$ since it underestimates heating of stellar kinematics, due to neglect of nonlinear phenomena like central solitons \cite{Schive:2019rrw}. In \cref{sec:comparison}, we compare the heating rates found in full, nonlinear Schrödinger–Poisson simulations to the estimates from the perturbative method \cite{Dalal2021} used in DK22. We show that the constraints derived using the latter are indeed conservative.

A second reason to revisit this topic is recent suggestions that tidal stripping could somehow weaken or even invalidate these constraints \cite{Yang2025}. In \cref{sec:tides}, we investigate the suppression of heating in the relevant regime for the actual satellite galaxies used to constrain ULDM.

A third reason to revisit FDM heating is recent suggestions that DM self-interactions could significantly alter the bounds on the DM particle mass derived from ultra-faint dwarf galaxies \cite{Capanelli2025}. We study this question in \cref{sec:selfinteraction}.

And finally, by far the most compelling reason to revisit bounds on ULDM is that galaxies even smaller than Segue 1 and Segue 2 may have been discovered \cite{Smith2024, Simon2024}. These objects, called ``micro-galaxies'' \cite{Errani2024}, have sizes more than an order of magnitude smaller than previous ultra-faint dwarf galaxies, and therefore should provide even stronger bounds on ULDM if they are confirmed as galaxies dominated by DM. These tighter bounds can have significant impact on laboratory searches for ultra-light axion DM \cite{Oshima2023, Bourhill2023, Shaw2022, Schulthess2022, Abel2017, Fan2024, Badurina2025}. In \cref{sec:UMa3}, we provide updated constraints on the DM particle mass using the best-studied micro-galaxy candidate, named Ursa Major~III.

\section{Simulation comparison}
\label{sec:comparison}

\begin{figure}
    \centering
    \includegraphics[width=\linewidth]{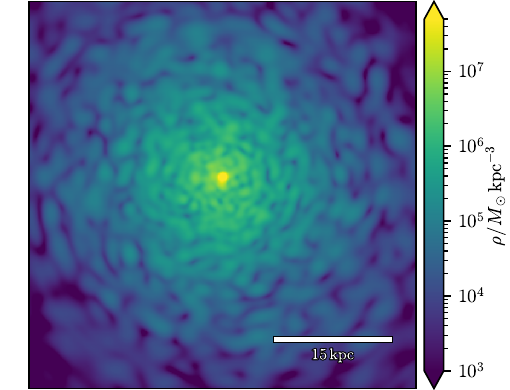}

    \caption{Projected density in a slab of thickness \SI{5}{\kpc} through the halo simulated using \texttt{AxiREPO}. See text for more details.}
    \label{fig:halo-projection}
\end{figure}

\begin{figure}
    \centering
    \includegraphics[width=\linewidth]{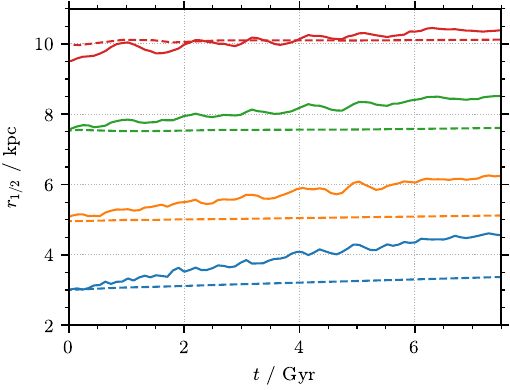}

    \caption{Evolution of the half-light radius of test particle groups in full nonlinear SP simulations (solid lines) and with the perturbative method (dashed lines).}
    \label{fig:comparison}
\end{figure}

In this section, we compare the perturbative method used in DK22 against full nonlinear Schrödinger–Poisson simulations. We use the \texttt{AxiREPO} code \cite{May2021, May2023} to evolve the DM wavefunction under the combined Schrödinger and Poisson equations. Because the stellar mass in ultra-faint dwarfs like Segue~1 comprises $\approx 10^{-3}$ of the observed gravitating mass within the stellar half-light radius \cite{Simon2011}, we neglect stellar self-gravity and model stellar motions within this halo using massless test particles. To make the simulations relatively inexpensive, we simulate a very light particle mass $m = \SI{e-22}{\eV}$. Although this mass is ruled out by observation, we nonetheless can make use of these simulations because we are only comparing the relative strength of FDM heating between perturbative simulations and full nonlinear simulations. We do not expect results of this comparison to change with $m$ as long as we scale units appropriately, i.\,e.\ compare heating rates at radii that are scaled with the de Broglie wavelength $\lambda_{\mathrm{dB}} = h/(m\sigma_v)$ (with Planck's constant $h$).

We simulate a cubic volume of side length \SI{150}{\kpc}, in a box with $960^3$ spatial Cartesian grid cells.
The spectral method employed by \texttt{AxiREPO} implies periodic boundary conditions, so in order to avoid spurious effects from the halo's periodic images, the simulation box size was chosen much larger than the size of the simulated halo.
Additionally, we further avoid periodic interference by placing an absorbing ``sponge'' region on the outer simulation boundary via the imaginary potential method \cite{2004PhRvD..69l4033G, 2016PhRvD..94d3513S}.

We generated an isolated FDM halo by colliding solitons (e.\,g.\ \cite{2016PhRvD..94d3513S}), to produce a halo with a central soliton whose outer density profile after relaxation resembles an NFW profile \cite{1997ApJ...490..493N} with a characteristic density and scale radius of $\rho_{\mathrm{s}} = \SI{4.8e6}{\Msun\per\kpc\cubed}$ and $r_{\mathrm{s}} = \SI{4.2}{\kpc}$. We evolved this halo for many \si{Gyr}s, measure the mean (time-averaged) density profile, and generated sets of massless test particles of various different initial half-light radii $\rh$, with velocity dispersions set according to the Jeans equation applied to the mean potential. As in DK22, the particles within each set have an initial (number) density profile that corresponds to an exponentially decreasing projected surface density profile. We then evolved these sets of test particles in the self-consistently evolved halo, which includes the central soliton and wave interference ``granules'', as can be seen in \cref{fig:halo-projection}.

We next constructed a comparison halo with the same mean profile using the Widrow–Kaiser \cite{Widrow:1993qq} method, and evolved the exact same sets of test particles in this comparison halo using the perturbative method \cite{Dalal2021, Dalal2022}. In both cases, we evolved the simulations for some time before making the comparison to allow the particles to settle into true equilibrium after an initial transient period. \Cref{fig:comparison} shows the comparison between heating rates in these two types of simulations, across the range of $\rh/\lambda_{\mathrm{dB}}$ relevant for UFD constraints. We find in \emph{all} cases that FDM heating is stronger in full nonlinear simulations than in perturbative simulations. This result is expected, since the perturbative simulations do not capture an important heating source, namely the fluctuations of a dense central soliton. Across the range of $\rh/\lambda_{\mathrm{dB}}$ that are important for UFD constraints on FDM, solitonic heating appears to dominate over diffusive heating from interference ``granules'', in some cases by large factors.

The fact that the heating rate in the perturbative simulation is always smaller than the heating rate in the full nonlinear simulation across all relevant radii means that the perturbative method provides a conservative lower limit on FDM heating, and therefore a conservative lower limit on the DM particle mass, as argued by DK22.

\section{Tidal stripping}
\label{sec:tides}

\begin{figure}
    \centering
    \includegraphics[width=0.95\linewidth]{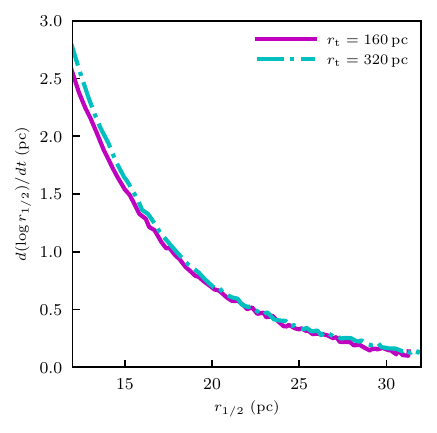}

    \caption{FDM heating rates for different tidal truncation radii. The two curves show the growth in half-light radii \rh\ for sets of test particles evolving inside of fluctuating FDM halos constructed to be consistent on average with the mass density observed within the half-light radius of Segue 1. Both simulations are for DM particle mass $m = \SI{2e-19}{\eV}$. The purple curve corresponds to a tidal radius $r_t=160$ pc, which is the smallest possible tidal radius given the mass observed within \rh\ of Segue 1, while the green curve corresponds to a tidal radius twice as large. As expected, tidal stripping produces no significant effect on heating rates, as long as the tidal radius is large compared to the size of the galaxy, as is observed for UFDs like Segue 1.}
    \label{fig:tide}
\end{figure}

A recent paper \cite{Yang2025} has argued that tidal stripping can weaken, or even invalidate, the bounds on ultra-light dark matter imposed by DK22.\footnote{Oddly, the simulation results shown in \cite{Yang2025} exhibit no evidence of tidal suppression of FDM heating, as seen in their Fig.~4. The central stellar densities of their simulations, both with and without external tides, are all depleted by about the same factors compared to the initial central densities. It appears that their simulation with the strongest tidal field gives a smaller half-light radius mainly because the tidal radius is similar in size to the half-light radius, as seen in their Fig.~3.}
The concern raised by \cite{Yang2025} is that ultra-faint dwarfs like Segue 1 are found within the Milky Way halo and, as satellites, they are subject to tidal gravitational fields that can strip away their dark matter. In principle, this stripping could suppress FDM heating of these satellites, when the tidal radius becomes as small as the soliton size \cite{Schive2020}.

It is quite easy to see that this argument is incorrect for observed UFDs like Segue~1, for the very simple reason that their tidal radii are much larger than the sizes of their stellar distributions \cite{Simon2019}. In this regime, the heating remains significant and is not sensitive to exact value of the tidal radius. As an illustration, \cref{fig:tide} shows results for the heating rates measured in two perturbative FDM simulations, in which groups of test particles of various half-light radii are evolved in FDM halos with particle mass $m = \SI{2e-19}{\eV}$. The two FDM halos are constructed using the Widrow–Kaiser method \cite{Dalal2021}, with mean density profiles that are truncated versions of the inner regions of NFW halos, with $\rho(r) \propto r^{-1} e^{-(r/r_{\mathrm{t}})^2/2}$, normalized to match the observed mass within the half-light radius of Segue~1 \cite{Simon2011}. Two different choices of the truncation radius $r_{\mathrm{t}}$ were used, either \SI{160}{\pc} (the minimum possible tidal radius for Segue~1, using only the mass interior to the half-light radius), and double that value. As the figure illustrates, changing the tidal radius by a factor of two produces no significant difference in the heating rates.

This result is completely unsurprising because these truncation radii greatly exceed the galaxy size and the de Broglie wavelength. As noted above, the origin of density fluctuations in FDM halos is wave interference. As long as there exist multiple waves of different frequencies and wavelengths, but comparable amplitudes, then there will necessarily be $\mathcal{O}(1)$ fluctuations in the resulting density field. Tidal stripping can suppress these fluctuations only when there is one single bound state in a halo whose amplitude is much larger than all other states, i.\,e.\ when the tidal radius becomes as small as the coherence length $\hbar/m\sigma_v$. In the mass regime of interest ($m > \SI{e-19}{\eV}$), the coherence length is much smaller than the half-light radius, which itself is much smaller than the tidal radius. So, quite obviously, wave interference is not significantly suppressed in this regime.

\begin{figure}
    \centering
    \includegraphics[width=0.95\linewidth]{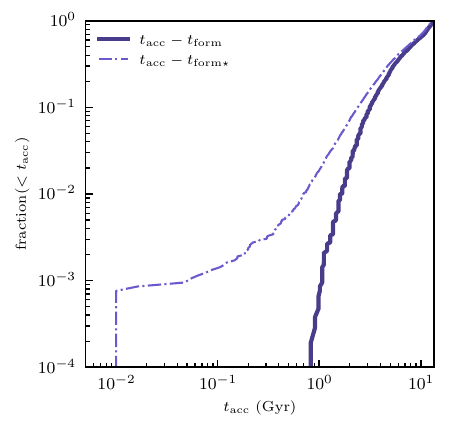}

    \caption{The cumulative distribution function of the accretion times of dark matter halos onto Milky Way-sized host halos in the Caterpillar simulation suite. The accretion times are measured relative to the time $t_{\mathrm{form}}$ when halo is first detected in the simulation (solid line) or relative to the time $t_{\mathrm{form}\star}$ when \SI{90}{\percent} of the stars of the galaxy it hosts have formed (dot-dashed line). The latter is estimated using a galaxy formation model described in the text.}
    \label{fig:tacc_cdf}
\end{figure}

At much lower masses ($m \lesssim \SI{2e-21}{\eV}$), the coherence length could become as large as the tidal radius, but as noted in DK22, those lower masses are already ruled out by a variety of observations (e.\,g.\ \cite{Nadler2021}). First, note that Segue~1 is not the only ultra-faint dwarf known. There exist many other UFDs, some with velocity dispersions and tidal radii much larger than that of Segue~1 \cite{Simon2019}. One example is DDO~210 \cite{Iorio2017, Hermosa2020}, which appears to be the central galaxy in an isolated halo, i.\,e.\ it is not a tidally stripped satellite. As noted by DK22, this galaxy already excludes masses below \SI{2e-21}{\eV}, and many other UFDs are observed spanning a wide range of sizes, velocities, and tidal radii that collectively exclude all masses below $m < \SI{3e-19}{\eV}$, also explained in DK22.

Secondly, and even more obviously, \emph{all} masses below $m < \SI{e-19}{\eV}$ are already ruled out by Segue 1 itself, including low masses (like $m \lesssim \SI{2e-21}{\eV}$) where the de Broglie wavelength is similar in size to the tidal radius. Even if Segue 1 were completely tidally stripped down to a bare soliton today, it nevertheless spent an enormous amount of time as the central galaxy in its own (unstripped) halo, before entering the Milky Way halo. At low masses, $m \lesssim \SI{2e-21}{\eV}$, FDM heating is orders of magnitude faster than at masses exceeding \SI{e-19}{\eV}. As noted in DK22, if Segue~1 were embedded in a soliton whose size exceeded the galaxy's size, then soliton fluctuations alone would destroy the galaxy on a timescale of order \num{e8} years, using the measured mean density and dynamical time within Segue~1. To avoid this, Segue~1 would need to lose its halo on a timescale faster than \num{e8} years.

Simulations of cosmological structure formation show that a typical subhalo in a Milky Way-sized halo that survives to $z = 0$ spends several billion years as an isolated halo prior to accretion (e.\,g.\ \cite{dsouza_bell21}). Indeed, \cref{fig:tacc_cdf} shows that the fraction of halos that spent $<10^8$ years as isolated halos before they were accreted is $<10^{-4}$ (see solid line) in the Caterpillar suite of simulations of Milky Way-sized halos and their satellites \cite{Griffen2016}. The fraction remains small, even if we take into account the fact that the relevant time scale is the time between the formation of most of the stars in a halo and the time of its accretion onto the Milky Way-sized host halo (the dot-dashed line in \cref{fig:tacc_cdf}). To estimate the formation time of the stars we use the galaxy formation model of \cite{Kravtsov2021} which was shown to reproduce the main statistical properties of ultra-faint dwarf galaxies \cite{Kravtsov2021, Manwadkar2022}.
Note also that it would take at least an additional orbital period following accretion, or $\approx \SI{1}{\Gyr}$ for a typical satellite galaxy \cite{Simon2018}, before any stripping could occur. Thus, dwarf galaxies would be exposed to dynamical heating by wave interference patterns for at least $\approx \SIrange{1}{2}{\Gyr}$, and typically for longer periods of time. Segue~1 would therefore have to be a $> \SI{99}{\percent}$ outlier in order to remain as small as it is today, if the DM mass were $m \lesssim \SI{2e-21}{\eV}$. And even in that case, as noted above, such DM masses are ruled out anyway by other dwarf galaxies.

Therefore, on multiple grounds, it is quite clear that tidal stripping cannot alter the constraint imposed on DM masses by ultra-faint dwarf galaxies.

\section{Self-interactions}
\label{sec:selfinteraction}

Previously-derived bounds on the DM mass $m$ \cite{Dalal2022} assume that non-gravitational interactions of DM particles are negligible compared to gravitational forces. In principle, DM self-interactions could alter FDM heating \cite{Capanelli2025}, and therefore could potentially change these constraints.

The argument for this scenario is as follows. FDM heating depends only on the DM density $\rho$, the fluctuation coherence length $\hbar/(mv)$, and the fluctuation coherence time $\hbar/(mv^2)$. In order to change the FDM heating rate, self-interactions must change one or more of these quantities. The mean DM density in dwarf galaxies is directly constrained by observed stellar kinematics, so self-interactions cannot change that ingredient. To change the constraint on $m$, therefore, self-interactions must change the DM velocity $v$ in observed dwarf galaxies. 

Previous analyses like DK22 assume that the DM velocity is determined by virial equilibrium, i.\,e.\ the velocity dispersion equals the value required to support the DM against self-gravity.
However, if DM self-interactions are significant compared to self-gravity, the equilibrium DM velocity dispersion can be significantly different than the virial velocity. In particular, if self-interactions are attractive, a larger velocity dispersion is required to support the DM against the combination of attractive gravity and attractive self-interaction. To change bounds on the DM mass appreciably, attractive self-interactions must be significantly stronger than self-gravity.

It is easy to see how strong a self-interaction is required to be to dominate over self-gravity. If the DM interaction potential is written (in natural units where $\hbar=c=1$) as
\begin{equation} \label{eq:V}
    V(\phi) = \frac{1}{2} m^2 \phi^2 + \frac{\lambda}{4} \phi^4 + \ldots
\end{equation}
then the Schrödinger equation for the complex field $\psi$ (``wavefunction'') , defined from the real scalar DM field via $\phi=(2m)^{-1/2} \left(e^{-imt}\psi + e^{imt}\psi^* \right)$, becomes 
\begin{equation}
    \label{eq:schrodinger}
    i  \frac{\partial\psi}{\partial t} \approx -\frac{1}{2m} \nabla^2 \psi
     + \frac{3\lambda}{4m^2} |\psi|^2 \psi + m \Phi \psi + \ldots,
\end{equation}
where $\Phi$ is the Newtonian gravitational potential, and note that the local density of DM is $\rho=m|\psi|^2$. Therefore, in order for the self-interaction term $3\lambda \rho/(4m^3)$ to dominate over the gravitational term $m\Phi$, we require a quartic coupling constant satisfying
\begin{equation}
    \label{eq:lower}
    |\lambda| > \frac{4\,m^4 c\,\Phi}{3\,\hbar^3\rho}.
\end{equation}
A similar requirement on $\lambda$ for self-interactions to dominate over gravity is given in Ref.~\cite{Capanelli2025}. Note that in \cref{eq:lower}, we have restored factors of $\hbar$ and $c$, to aid in numerical evaluation. 

The difficulty in satisfying this requirement on the coupling is that the same self-interaction is also present in the early universe, where the mean DM density is just as large as (or significantly larger than) the density in DM halos at $z=0$. Therefore, if self-interactions dominate over self-gravity today in the $z=0$ DM halos, they can also dominate over self-gravity in the early universe, where they would create large differences in the behavior of DM compared to standard $\Lambda$CDM.

\begin{figure}
    \centering
    \includegraphics[width=\linewidth]{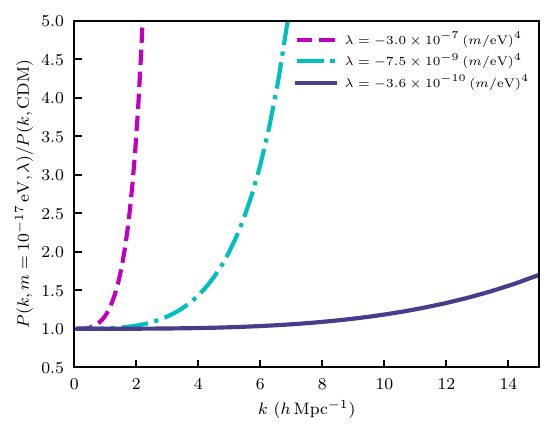}

    \caption{The effect of DM self-interactions on the linear power spectrum. Plotted are two examples of the ratio of the linear power spectrum for $m = \SI{e-17}{\eV}$ and nonzero quartic coupling $\lambda$, compared to the standard $\Lambda$CDM power spectrum. These curves are computed by solving \cref{eq:linearevol}, derived by \cite{Arvanitaki2015}.}
    \label{fig:power}
\end{figure}

One example is that strong self-interactions in the early universe can generate $\mathcal{O}(1)$ changes to the DM power spectrum on observable scales. 
We can estimate the growth of linear perturbations for DM with ultra-light mass $m$ and quartic coupling $\lambda$ by solving the linearized evolution equation for the overdensity perturbation $\delta_{\mathbf{k}} = \delta\rho_{\mathbf{k}}/\bar{\rho}_{\mathrm{m}}$ at comoving wavenumber $\mathbf{k}$ \cite{Arvanitaki2015},
\begin{equation}
    \label{eq:linearevol}
    \ddot{\delta}_{\mathbf{k}} + 2 H \dot{\delta}_{\mathbf{k}} \approx
    \left[4\pi G \bar{\rho}_{\mathrm{m}} - \frac{k^4}{4m^2 a^4} -
    \frac{3\lambda\bar{\rho}_{\mathrm{m}} k^2}{8m^4 a^2}\right] \delta_{\mathbf{k}},
\end{equation}
where $G$ is Newton's constant, $H$ is the Hubble parameter, $a$ is the cosmic scale factor, and $\bar{\rho}_{\mathrm{m}}$ is the mean matter density.
Note that this linearized expression may not be applicable if density perturbations become highly nonlinear in the early universe, leading to the formation of extremely dense structures like oscillons. We do not consider that regime here, i.\,e.\ we assume that cosmic structure remains sufficiently linear in the early universe for self-interacting models to remain consistent with observed large-scale structure on linear and quasi-linear scales.

\Cref{fig:power} shows examples of how $\lambda$ changes the DM linear power spectrum as a function of comoving wavenumber $k$. Attractive self-interactions ($\lambda<0$) can enhance the small-scale power spectrum, on scales that are constrained by high-resolution observations of the Lyman-$\alpha$ forest \cite{Irsic2024} or the abundance of Local Group satellites \cite{Nadler2021}. If we impose the constraint that the power spectrum cannot deviate from the $\Lambda$CDM power spectrum by more than a factor of 10 for $k\leq \SI{30}{\hHubble\per\Mpc}$, as indicated by the central densities of observed ultra-faint Milky Way dwarf galaxies \cite{Esteban2024, Dekker2025}, then the quartic coupling is bounded from above by
\begin{equation}
    \label{eq:upper}
    |\lambda| < \num{3.6e-10} \left(\frac{m}{\si{\eV}}\right)^4.
\end{equation}

We can immediately see that it may not be possible to satisfy both the lower limit in \cref{eq:lower} and the upper limit in \cref{eq:upper}. If we estimate the characteristic density in  \cref{eq:lower} using virial equilibrium as $\rho\approx 9\sigma_v^2/(4\pi G \rh^2)$, and estimate the characteristic potential depth as $\Phi \approx \sigma_v^2$ (which is an underestimate if $\sigma_v$ is the stellar velocity dispersion), then we see that there is no $\lambda$ satisfying both \cref{eq:lower} and \cref{eq:upper} if $\rh > \SI{1.1}{\pc}$. All known galaxies have \rh\ exceeding this size, including Segue~1, Segue~2, and all of the other galaxies used to constrain the DM particle mass \cite{Dalal2022}, which means that self-interactions cannot change the constraints imposed by these galaxies without simultaneously also violating bounds from the linear power spectrum.

Note that we have only considered constraints from the linear regime of structure formation. In the nonlinear regime, strong self-interactions generate a plethora of additional effects, such us the formation and decay of dense structures called oscillons \cite{Arvanitaki2020}, which impose additional constraints on $\lambda$ beyond \cref{eq:upper}. We have not considered these additional nonlinear effects here, nor have we considered effects of higher-order terms in the potential beyond the terms in \cref{eq:V}, but we note that any proposal for ULDM that attempts to evade the bounds from galaxy kinematics must first demonstrate that it is consistent with constraints on DM interactions from structure formation in both the linear and nonlinear regimes.

In summary, it appears that DM self-interactions do not change the constraints imposed by dwarf galaxy kinematics, as long as those self-interactions are consistent with large-scale structure.

\section{Ursa Major III/UNIONS 1}
\label{sec:UMa3}

The previous sections showed that a conservative estimate of FDM heating may be obtained using the perturbative approach of \cite{Dalal2021}, that this estimate is insensitive to tidal stripping of satellites as long as the tidal radius is large compared to the satellite size and de Broglie wavelength, and that the constraint on heating is not affected by DM self-interactions within the range of interaction strengths allowed by existing constraints on the matter power spectrum. This perturbative method was previously used to derive lower limits on the DM particle mass, using the smallest known ultra-faint dwarf galaxies Segue~1 and Segue~2 \cite{Dalal2022}. These bounds are the most stringent existing bounds on the mass of the DM particle. However, they may be superseded, thanks to the possible discovery of a class of even smaller galaxies, termed ``micro-galaxies''. The prototype for this class is the object Ursa Major~III/\allowbreak UNIONS~1, which has a projected 2D half-light radius of $r_{\mathrm{h}} = \SI{3 +- 1}{\pc}$ \cite{Smith2024}, roughly corresponding to a 3D half-light radius of $\rh \approx \SI{4 +- 1.3}{\pc}$. The measured velocity dispersion $\sigma_v = \SI{1.9 +- 1.4}{\km\per\s}$ is much larger than the velocity required to support this object against self-gravity from its stellar mass of $M_* \approx \SI{16}{\Msun}$, suggesting that this object is highly dark matter-dominated with $M(<\rh) \approx 3\sigma_v^2\rh/G \approx \SI{e4}{\Msun} \gg M_*$. This dispersion measurement is still somewhat uncertain, however, since it is quite sensitive to outlier rejection, and it relies on single-epoch spectroscopy, meaning that binary contamination could artificially inflate the measured dispersion.

\begin{figure}
    \centering
    \includegraphics[width=0.95\linewidth]{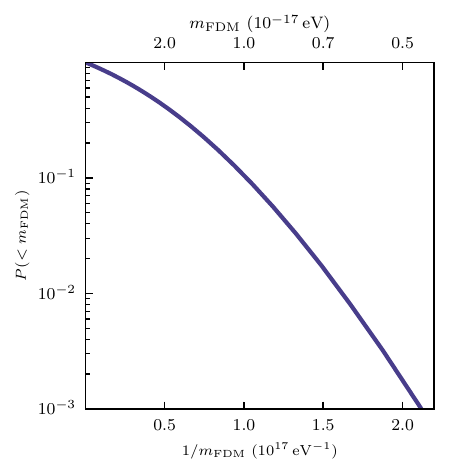}
    \caption{Cumulative posterior on DM mass $m$ from Ursa Major III, using $\rh = \SI{4 +- 1.3}{\pc}$ in 3D, and $\sigma_v = \SI{1.9 +- 1.4}{\km\per\s}$ \cite{Smith2024}. We find that $m > \SI{8e-18}{\eV}$ at \SI{95}{\percent} confidence.}
    \label{fig:bounds}
\end{figure}

In spite of these uncertainties, there are many indications that Ursa Major III is a galaxy and its mass is dominated by dark matter. One such indication is its survival.  Its stellar mass is only \SI{16(6:5)}{\Msun}, while the age of its stellar population is $\gtrsim \SI{10}{\Gyr}$ \cite{Smith2024}. As shown by \cite{Errani2024}, a star cluster of such mass would be completely disrupted by tidal forces on a timescale of order 400 Myr, much shorter than the ages of its stars.  This would require us to observe this system at a special time in its history, in its final orbit around the Milky Way.  Since this object is not on its first orbit, then for it to be a star cluster it must have had a much larger mass in the past, to allow it to survive in the Milky Way tidal field for many orbits.

A recent paper \cite{Devlin2025} studies exactly this scenario, exploring the possibility that Ursa Major~III/\allowbreak UNIONS~1 could be the remnant of an initially much more massive star cluster.  This work argues that retention of stellar remnants and their mass segregation would allow a star cluster with the observed properties of Ursa Major~III to survive for $\approx \SIrange{1}{2}{\Gyr}$, a few times longer than the estimate of \cite{Errani2024}, but still much shorter than the age of the system's stars.  This scenario must overcome significant challenges to be feasible.  First, in the models of \cite{Devlin2025}, the star cluster initially contains $\approx \mathcal{O}(6000)$ stars, but it loses almost all of its stars over time, leaving behind just a few dozen stars seen today. This enormous number of lost stars would presumably lead to a prominent stellar stream tracing the orbit of Ursa Major~III.  For example, the progenitor of the GD-1 stream is believed to have a mass $M\sim 10^4 M_\odot$ \cite{Webb19a,Gialluca2021}, similar to the initial cluster mass used in \cite{Devlin2025}, and the GD-1 orbital period, pericentre and apocentre are roughly similar to the corresponding parameters for Ursa Major III.  That stream was detected decades ago, using (by modern standards) shallow photometry of SDSS \cite{GD1} however no such stellar stream is observed by UNIONS along the orbit of Ursa Major III \cite{Smith2024}. This problem is compounded by the fact that Ursa Major III was not formed in the Milky Way, but instead was accreted. This is known from the inferred orbital parameters of this system, with a pericenter of \SI{12.6 +- 0.7}{\kpc} and an apocenter of \SI{26(6:7)}{\kpc} \cite{Smith2024}, which are inconsistent with the orbital distances of surviving in-situ star clusters born in the Milky Way, typically smaller than \SI{10}{\kpc} \cite{Belokurov2024}.
Therefore, if Ursa Major~III were a star cluster, it must have been accreted as part of a galaxy. That galaxy could thus also contribute to the stellar stream around the cluster, which was not detected. Also, star clusters are born in the dense interstellar medium of their parent galaxies and they experience strong tidal forces from their birth. However, the simulations in \cite{Devlin2025} account only for the tidal forces of the Milky Way, but not for the tidal forces the cluster would experience in its natal environment during early stages of its evolution. The tidal mass loss of the cluster and its dissolution time in their simulations may therefore be under-/over-estimated, respectively. 

The weight of existing evidence thus suggests that Ursa Major~III is indeed a dwarf galaxy dominated by dark matter, which would make it the smallest known galaxy (but note that other candidate micro-galaxies exist \cite{Simon2024}). As discussed above, the smallest galaxies provide the most sensitive probes of ULDM, so we can expect Ursa Major~III to give the strongest lower limit on the mass of the DM particle. Using the back-of-the-envelope estimate described in \cite{Dalal2022}, we expect that this galaxy should provide a limit of $m \gtrsim \SI{e-17}{\eV}$.

In principle, we could simulate the evolution of this system inside FDM halos of various particle mass for \SI{10}{\Gyr}, but for $m = \SI{e-17}{\eV}$ the expected coherence time is at most $\hbar/(m\sigma_v^2) \approx \SI{5e4}{\yr}$. If we resolve the coherence time with at least five time steps, then each simulation would require at least $10^6$ time steps to evolve for \SI{10}{\Gyr}. This simulation suite would be expensive, and quite tedious, so we instead adopt a simpler approach. Using shorter simulations, we measure the growth rate of the half-mass radius of sets of test particles, $\dl\rh/\dl t$, evolved inside of FDM halos, as a function of the DM particle mass $m$, mean halo parameters, and \rh. We then integrate $\dl\rh/\dl t$ to determine the time $T$ required to grow to a final size of \rh, starting from an initial size that is much smaller than the coherence length, for each different halo potential and each different $m$. If this time $T$ is smaller than the observed age of the galaxy (taken to be \SI{10}{\Gyr}), then this galaxy cannot remain as small as observed in that potential with that DM mass $m$.

We assume a halo potential given by the deep interior region of an NFW halo, with $\rho(r) \propto r^{-1}$, and we use the circular velocity $v_{\mathrm{c}}(r_{\mathrm{fid}})$ as the parameter setting the normalization of the halo potential, for fixed fiducial radius $r_{\mathrm{fid}} = \SI{4}{\pc}$. For this simple potential, the virial theorem gives us a simple relation between the stellar velocity dispersion and the stellar half-light radius, $\sigma_v^2 \approx v_{\mathrm{c}}^2(r_{\mathrm{fid}}) \rh/(3r_{\mathrm{fid}}$). We then use the observed properties of Ursa Major~III, namely $\rh = \SI{4 +- 1.3}{\pc}$ and $\sigma_v = \SI{1.9 +- 1.4}{\km\per\s}$ \cite{Smith2024}, as Gaussian priors on \rh\ and $\sigma_v$ to derive a likelihood for the mass $m$,
\begin{widetext}
\begin{equation}
    \mathcal{L}(m) = \int_0^\infty \dl\rh \, P_r(\rh) 
    \int_{\sigma_{\mathrm{min}}}^\infty \dl\sigma_v \,
    P_\sigma(\sigma_v)\, 
    \Theta\mleft(T(m, \rh, v_{\mathrm{c}}) - \SI{10}{\Gyr}\mright),
\end{equation}
\end{widetext}
where $\Theta$ is the Heaviside step function, $T(m, \rh, v_{\mathrm{c}})$ is the time required to grow to size \rh\ for DM particle mass $m$ in a potential with normalization $v_{\mathrm{c}}(r_{\mathrm{fid}}) = \sigma_v \sqrt{3r_{\mathrm{fid}}/\rh}$, $P_r$ is the prior on \rh\ which we assume to be a Gaussian with mean of \SI{4}{\pc} and root mean square of \SI{1.3}{\pc}, and $P_\sigma$ is similarly the prior on the stellar $\sigma_v$, taken to be a Gaussian with mean \SI{1.9}{\km\per\s} and root mean square \SI{1.4}{\km\per\s}. We neglect covariance between the uncertainties on \rh\ and $\sigma_v$.  Following DK22, we define the posterior probability as the product of this likelihood and our prior on $m$, which we assume to be $P\propto m^{-2}$, corresponding to a uniform prior on $m^{-1}$. \Cref{fig:bounds} shows the resulting cumulative probability that the DM mass is smaller than $m$. We find that $m > \SI{8e-18}{\eV}$ at \SI{95}{\percent} confidence, in excellent agreement with the back-of-the-envelope estimate.

Note that this lower bound is quite conservative. First, our assumed $m^{-2}$ prior strongly favors light masses over heavy masses, weakening the lower limit. Secondly, as we saw in \cref{sec:comparison}, the perturbative method used to derive this bound gives heating rates significantly smaller than the heating found in full nonlinear simulations. We expect that the bound on $m$ could be considerably strengthened using full Schrödinger–Poisson simulations. Note that, as mentioned previously, the stellar mass comprises a fraction $\approx \mathcal{O}(\num{e-3})$ of the mass interior to the half-light radius, meaning that the neglect of stellar self-gravity produces a negligible error in these simulations.

\section{Discussion}

In this paper, we have presented updated lower bounds on the mass of the DM particle, using the kinematics of the smallest galaxies. We find that $m > \SI{8e-18}{\eV}$ at \SI{95}{\percent} confidence, under the assumption that Ursa Major III / UNIONS 1 is confirmed as a galaxy dominated by dark matter. This constraint is by far the strongest lower limit on $m$, and in comparison with other constraints such as satellite counts \cite{Nadler2021} or the Lyman-$\alpha$ forest \cite{Hui2021}, this bound is completely insensitive to uncertainties in baryonic physics such as feedback from star formation, the thermal history of the intergalactic medium, the galaxy–halo connection, etc. The reason for this robustness is that we do not rely on predictions for the central DM density in these dwarf galaxies, but instead make use of the empirically observed DM densities. Because these objects are so thoroughly dominated by dark matter, with $M_{\mathrm{DM}}/M_* \approx 600$ inside the stellar half-light radius, the dynamics are particularly simple to model. 

The modeling is also simplified by our results in \cref{sec:tides} showing that FDM heating in ultra-faint dwarfs should not be affected by tidal stripping given lower limits on the tidal radii of these galaxies, and by our results in \cref{sec:selfinteraction} showing that FDM heating in UFDs cannot be reduced by DM self-interaction within the range of self-interaction strength allowed by observations. As shown in \cref{sec:comparison}, our constraints are quite conservative since they do not account for the heating by the fluctuations of the central soliton, which can dominate the overall dynamical heating, and thus certainly underestimate the effect on the stellar system (see \cref{fig:comparison}). Our constraint on the DM particle mass could therefore be significantly strengthened in the future if full nonlinear simulations are used to estimate the heating, instead of the more conservative perturbative method that we have adopted.

The constraint we obtain means that if dark matter is indeed composed of ultra-light particles, it is indistinguishable from cold dark matter using all existing probes of small-scale structure. As already noted in DK22, any deviations in the linear power spectrum that can arise in ULDM (compared to CDM) arise at wavenumbers that cannot be probed using quasi-linear signals like the Lyman-$\alpha$ forest, or nonlinear probes like satellite counts, strong gravitational lensing or tidal streams. Our new, more stringent bounds do not alter that conclusion. For $m > \SI{8e-18}{\eV}$, the linear power spectrum will be similar to $\Lambda$CDM power spectrum for $k < \SI{e3}{\hHubble\per\Mpc}$ \cite{Hu:2000ke}, and the halo mass function will match the $\Lambda$CDM halo mass function for masses $M \gtrsim \SI{e3}{\per\hHubble\Msun}$, scales far beyond the reach of any currently known cosmological probes. Detecting wave interference using gravitational lensing remains utterly hopeless in light of these bounds, since typical strong lenses are massive galaxies with circular velocities of order \SI{300}{\km\per\s}, meaning that the coherence length would be of order $\hbar/(mv) < \SI{e-3}{\pc}$, so that $\mathcal{O}(1)$ density fluctuations at Einstein radii of order \SI{5}{\kpc} will correspond to $< \mathcal{O}(\num{e-3})$ fluctuations in the projected convergence.

One area where our bound may have an impact is on terrestrial experiments searching for evidence of ULDM via its non-gravitational interactions with ordinary matter, e.\,g.\ \cite{Oshima2023, Bourhill2023, Shaw2022, Schulthess2022, Abel2017, Fan2024, Badurina2025}. Searches for dark matter at masses $m < \SI{e-17}{\eV}$ thus do not appear to be worthwhile to pursue if Ursa Major~III is a galaxy dominated by dark matter.

\acknowledgments
We thank Mina Arvanitaki, Ana Bonaca, Scot Devlin, Raphael Errani, Junwu Huang, Norm Murray, Adrian Price-Whelan, Ken van Tilburg, and Claire Ye for helpful discussions.
ND is supported by the Centre for the Universe at Perimeter Institute, and by MEXT KAKENHI Grant Number JP20H05861.
Research at Perimeter Institute is supported in part by the Government of Canada through the Department of Innovation, Science and Economic Development Canada and by the Province of Ontario through the Ministry of Colleges and Universities. AK was supported by the NASA ATP grant 80NSSC20K0512 and the National Science Foundation grant AST-2408267.
This research was enabled in part by resources provided by Compute Ontario and the Digital Research Alliance of Canada, and computational resources provided by the University of Chicago Research Computing Center. 
This work has made use of the \texttt{GSL} \cite{GSL}, \texttt{FFTW} \cite{FFTW}, \texttt{SHTns} \cite{shtns}, \texttt{NumPy} \citep{numpy_ndarray}, \texttt{SciPy} \citep{scipy}, and \texttt{Matplotlib} \citep{matplotlib} libraries, and we thank the respective authors for making their software publicly available. 
We have also used the Astrophysics Data Service (\href{https://adsabs.harvard.edu/}{\texttt{ADS}}) and \href{https://arxiv.org}{\texttt{arXiv}} preprint repository extensively during this project and the writing of the paper.

\bibliography{fdm}

\end{document}